# Optimum window length of Savitzky-Golay filters with arbitrary order

Mohammad Sadeghi, Fereidoon Behnia

Department of Electrical Engineering, Sharif University of Technology, Tehran, Iran. Emails:sadeqi.muhamad@gmail.com, behnia@sharif.

**Abstract**

One of the widely used denoising methods in different domains is the Savitzky-Golay (SG) filter. The SG filter has two design parameters: window length and the filter order. As the length of the window increases, the estimation variance decreases, but the bias error increases at the same time .Mean square error (MSE) measure includes both bias and variance criteria. In this paper, we obtain the optimal window length of an SG filter with arbitrary order which minimizes the MSE. To achieve the optimal window length, we propose an algorithm whose performance is better than the existing methods. In this paper, we follow the viewpoint proposed by Persson and Strang and design the filter on the basis of Chebyshev orthogonal polynomials

## I. Introduction

In real world and practical applications, measured signals are polluted by noise. There are numerous methods in the literature for signal denoising such as wavelet denoising [1-2], total variation (TV) filtering [3-4], Stein's unbiased risk estimate (SURE) regularization [6-7], Savitizky-Golay and legendre filters [8-9], kernel regression [10-11] and denoising based on sparsity of the signal [12-13]. In this paper, we use Savitzky-Golay and legendre filter for denoising. This filter locally smooths a noisy signal by fitting a specified order polynomial to samples in the window in least square sense. Whenever the criteria for evaluating a filter are computational complexity, capability of denoising and waveform reconstruction, Savitizky-Golay filter is an appropriate candidate. This filter was first introduced by Savitzky and Golay in analytical chemistry as a solution for smoothing noisy data obtained from the chemical spectrum analyzer. This filter is currently used in extensive applications such as digital control systems [14], ridge detection in image processing[15], speech recognition[16], gas concentration detection based on tunable diode lasers [17],EEG signal enhancement[18], post seismic vegetation recovery [19], coronary wave analysis [20-21], online condition monitoring of power transformer on-load tap changer [22] ,spectrogram enhancement in bird sound detection [23],Electrocardiogram denoising [24-25] and reactivity calculation in nuclear power [26].

Despite widespread use of SG filter, few theoretical work has been done/reported on this subject. In the following, we review the theoretical work in this area. The characteristics of the SG filter are discussed in [27-29]. In [30-31] non-uniform sampling mode of this filter has been studied. The popular SG filter is an FIR filter. In [32] this filter is generalized to an IIR filter and compared with the FIR version of the filter. The estimation performance of SG filter was analyzed in [33]. The typical SG filter coefficients are optimal for Gaussian noise. For heavy tailed distribution of noise, optimal coefficients are investigated in [34]. In [35] in the smoothing application, the number of multipliers for calculating filter output was reduced.

In this paper, we use 1D FIR SG filter assuming uniform sampling for denoising and smoothing the noisy signal. The SG filter has two design parameters: window length and order of filter. The larger the window length, the less the filter output error variance, but whenever the window length is selected too high, the filter output becomes biased in comparison to the actual signal. Mean square error (MSE) measure has both bias and variance criteria.

We are looking for the optimal window length so that the distance between the estimated and the actual signal is minimized in MSE sense. Selecting the appropriate window length for kernel regression issues has been widely discussed ([36] and references in it), but the optimal window length of SG filter has not been considered sufficiently.

In [8], [37-38] the effect of the window length has been considered only numerically. In [39] using SURE theory, MSE cost function is converted to an appropriate cost function versus window length and then for different window lengths, the cost function is evaluated and the optimum window length is determined based on its minimum value. To reduce computational complexity of the algorithm, space search of the window length is limited by a preset maximum window length ($N_{max}$). The authors of [39] believe that $N_{max} = 64$ is sufficient for satisfactory performance in their experiments; but this claim is not exact, as we show in this paper that the optimal window length depends on the noise power, number of samples, order of filter and signal waveform. Our proposed algorithm has better performance compared to the method presented in [39] and at the same time has lower computational burden. It should be noted that the design of the SG filter in [39] is based on the polynomial fitting, while our method is based on the Chebyshev orthogonal polynomials viewpoint presented in [8] by Persson and Strang.

The rest of this paper is organized as follows. In section II the signal model and a review of SG filter are presented. SG filter design baes on Chebyshev orthogonal polynomials is described in section III. In section IV optimal window length for SG filter with arbitrary order is calculated. Numerical results are presented in section V and finally, concluding remarks are discussed in section VI.

## II. Signal model

Consider the main signal $f(t)$ corrupted by additive independent and identically distributed (iid) noise $w(t)$ with zero mean and variance $\sigma^2$. The noisy signal $x(t)$ can be modeled as:

$$x(l) = f(l) + w(l) \qquad l = 1, \dots, L$$

where $f(l)$ indicates the $l$th sample of $f(t)$. Having $L$ samples of noisy signal $x(t)$, our purpose is to retrieve $f(t)$ from these samples using an SG filter. In the following we review the SG filter based on [9],[39].

**A review of SG filter**

Consider a symmetric window with a length of $N = 2M + 1$ samples around the reconstruction point, then a polynomial with order of $n$ ($P(i) = \sum_{k=0}^{n} a_k i^k$  $k = 0, \dots, n$ where $a_k$ is the $k$th coefficient of the polynomial) is fitted to the samples within the window in such a way as to minimize the following MSE.

$$\varepsilon_n = \sum_{i=-M}^{M} (P(i) - x(i))^2 = \sum_{i=-M}^{M} \left( \sum_{k=0}^{n} a_k i^k - x(i) \right)^2$$

the order of the polynomial ($n$) is called the filter order.

The filter output is equal to the value of the polynomial in the central point ($y(0)$) meaning

$$y(0) = p(0) = a_0$$

To calculate the next point of the filter output, the window is shifted by one unit and the operation is repeated.

Savithky and Golay showed that this process of filtering is equivalent to convolving samples in windows with a fixed impulse response. With this description, the filter output can be written as follows

$$y(k) = \sum_{i=-M}^{M} w_i x(k-i)$$

Which means that the weighted average of the samples in the window is equal to the filter output. To determine the coefficients of the filter, we differentiate $\varepsilon_n$ with respect to the these coefficients and set the derivatives equal to zero leading to $n+1$ equations in $n+1$ unknowns as follows

$$\sum_{k=0}^{n} \left( \sum_{i=-M}^{M} i^{j+k} a_k \right) = \sum_{i=-M}^{M} i^j x(i) \qquad j = 0,1,\ldots,n \quad (1)$$

We define the polynomial coefficient vector $\boldsymbol{a} = [a_0, a_1, \ldots, a_n]^T$, input samples vector $\mathbf{x} = [x_{-M}, \ldots, x_{-1}, x_0, x_1, \ldots, x_M]^T$ (where $x_i = x(i)$) and matrix $A$

$$A = \begin{pmatrix} (-M)^0 & (-M)^1 & \cdots & (-M)^n \\ \vdots & \vdots & \vdots & \vdots \\ (-1)^0 & (-1)^0 & \cdots & (-1)^0 \\ 1 & 0 & \cdots & 0 \\ M^0 & M^1 & \cdots & M^n \end{pmatrix}$$

to write equation (1) in matrix form as

$$(A^T A)^{-1} \boldsymbol{a} = A^T \mathbf{x}$$

With this notation, the coefficient vector can be derived as below

$$\boldsymbol{a} = (A^T A)^{-1} A^T \mathbf{x} = W\mathbf{x}$$

Note that we only need to compute $a_0$, so calculating the first row of the matrix $W$ is sufficient.

The matrix $W$ is independent of the input samples and only depends on the filter order ($n$) and the window length ($M$). Therefore the weighting coefficients are the same in all windows. Finally the filter output at index 0 is as the following

$$y(0) = \sum_{i=-M}^{M} w_{-i} x_i = a_0$$

Obtaining the optimal window length for an SG filter with order $n$ is very difficult from the polynomial fitting viewpoint. We can look at the SG filter from another point of view based on Chebyshev orthogonal polynomials. Persson and Strang have shown in [8] that SG filter can be designed using discrete orthogonal polynomials. This viewpoint is described in the next section based on [8].

### III. SG filter with Chebyshev orthogonal polynomials

The signal model is the same as the model presented in section II. Filter output with order $n$ and window length $N = 2M + 1$ can be written as follows

$$y(k) = \sum_{j=-M}^{M} W_n(j) x(k-j) = \sum_{j=-\frac{N-1}{2}}^{\frac{N-1}{2}} W_n(j) x(k-j)$$

Persson and Strang [8] have shown that the SG filter coefficients with order $n$ are samples of polynomials $W_n(x)$ in points $x = -M, \ldots, M$. The polynomial $W_n(x)$ is as follows

$$W_n(x) = \alpha_{n+1} \frac{q_{n+1}(x)}{x} \qquad (1)$$

where

$$\alpha_{n+1} = \frac{n+1}{2^{n+1}} \binom{n}{\frac{n}{2}} \frac{(-1)^{\frac{n}{2}}}{N(N^2 - 2^2)(N^2 - 4^2) \ldots (N^2 - n^2)} \qquad (2)$$

and $q_n(x)$ is a shifted Chebyshev polynomial that can be obtained by an $n$th forward difference $\Delta^n$

$$q_n(x) = n! \, \Delta^n \left[ \binom{x+M}{n} \binom{x-M-1}{n} \right] \qquad (3)$$

The first four polynomials are as follows

$$q_0(x) = 1$$
$$q_1(x) = 2x$$
$$q_2(x) = 6x^2 - 2M(M+1)$$

$$q_3(x) = 20x^3 - 4x(3M^2 + 3M - 1)$$

for example the coefficient of SG filter with order 2 can obtained as follows

$$W_2(x) = \frac{3}{8}(2)\frac{-1}{N(N^2 - 2^2)}\frac{q_3(x)}{x}$$

$$= \frac{3(3M^2 + 3M - 1)}{(2M + 1)(4M^2 + 4M - 3)}$$

$$+ \frac{-15}{(2M + 1)(4M^2 + 4M - 3)}x^2 \qquad x = -M, \ldots, M$$

The polynomials $q_n(x)$ have the following properties( [8] and [40])

**Property 1**: If $n$ is even, polynomial $q_n(x)$ is also even and vice versa.

Note: everywhere in this paper, we assume that $n$ is even.

**Property 2**: Polynomials $q_n(x)$ are orthogonal, that is

$$\sum_{x=-\frac{N-1}{2}}^{\frac{N-1}{2}} q_n(x)q_m(x) = 0 \qquad \forall m \neq n \qquad (4)$$

and that

$$\sum_{x=-\frac{N-1}{2}}^{\frac{N-1}{2}} q_n^2(x) = \frac{N(N^2 - 1^2)(N^2 - 2^2) \ldots (N^2 - n^2)}{2n + 1} \qquad (5)$$

**Property 3**: a recursive relation exists between the three consecutive polynomials $q_{n-1}(x)$, $q_n(x)$ and $q_{n+1}(x)$ as follows

$$(n + 1)q_{n+1}(x) = 2(2n + 1)xq_n(x) - n(N^2 - n^2)q_{n-1}(x) \qquad (6)$$

With $n$ even, $q_{n+1}(x)$ is an odd polynomial and so $q_{n+1}(0) = 0$ and $q_n(0)$ can be obtained following the recursive formula (6) as below.

$$q_n(0) = -\frac{(n-1)}{n}(N^2 - (n-1)^2)q_{n-2}(0) = \cdots$$

$$= \frac{(-1)^{\frac{n}{2}}}{2^n}\binom{n}{\frac{n}{2}}\sum_{k=1}^{\frac{n}{2}}(N^2 - (2k-1)^2)$$

$$(7)$$

The proof of these properties can be found in [8] and [40].

## IV. Optimum window length for SG filter with arbitrary order

In this section, we want to determine the optimum window length of SG filter with order $n$ based on the orthogonal polynomials and using the properties mentioned in section III.

To calculate the optimal window length, we need to know the first $p \in \mathbb{N}$ so that the statement $\sum_{i=-M}^{M} W_n(i) i^p$ is not zero. This $p$ is determined in Lemma1. Note that since $w_i = w_{-i}$ for odd $p$ the statement $\sum_{i=-M}^{M} W_n(i) i^p$ is always equal to zero.

**Lemma1**: if $W_n(i)$, $i = -M, \ldots, M$ are coefficient of an SG filter with order $n$ and

$$\sum_{i=-M}^{M} W_n(i) i^p = 0 \quad \forall p < p_0, \quad \{p, p_0\} \in \mathbb{N}$$

$$\sum_{i=-M}^{M} W_n(i) i^{p_0} \neq 0$$

then

$$p_0 = n + 2$$

**Proof**

It can be seen that the expression $x^p$ can be expressed in terms of the linear combination of Chetbyshev polynomials of order $p$ and lower, i.e. $x^p = \sum_{j=1}^{p} \tau_j q_j(x)$. Now we have

$$\forall p \leq n+1: \sum_{x} W_n(x) x^p = \alpha_{n+1} \sum_{x} q_{n+1}(x) x^{p-1}$$

$$= \alpha_{n+1} \sum_{x} q_{n+1}(x) \sum_{j=1}^{p-1} \tau_j q_j(x)$$

$$= 0$$

Note that the last equality is due to orthogonality of Chebyshev polynomials (Property 3). On the other hand, for p = n + 2 we have:

$$p = n + 2: \sum_x W_n(x)x^p = \alpha_{n+1} \sum_x q_{n+1}(x)x^{p-1} = \alpha_{n+1} \sum_x q_{n+1}(x)x^{n+1}$$

$$= \alpha_{n+1} \sum_x q_{n+1}(x) \sum_{j=1}^{n+1} \tau_j q_j(x) = 0 + \alpha_{n+1}\tau_{n+1} \sum_x q_{n+1}^2(x)$$

$$\neq 0$$

∎

As previously mentioned, our benchmark for optimal window length is minimum MSE. So the following problem should be solved.

$$\widehat{M} = \underset{k}{\operatorname{argmin}} E\left\{(y(t) - f(t))^2\right\}$$

$$= \underset{k}{\operatorname{argmin}} E\{y^2(t) - 2y(t)f(t) + f^2(t)\} \quad (8)$$

where $E$ denotes statistical expectation operator.

We compute the three terms in in (8). First, consider the first term

$$E\{y^2(t)\} = E\left\{\left(\left[\sum_{i=-M}^{M} W_n(i)\, x(t-i)\right]\left[\sum_{j=-M}^{M} W_n(j)\, x(t-j)\right]\right)\right\}$$

$$= \underbrace{E\left\{\left(\sum_i W_n(i)f(t-i)\right)^2\right\}}_{=\gamma}$$

$$+ \underbrace{E\left\{\sum_i \sum_j W_n(i)W_n(j)\, w(t-i)w(t-j)\right\}}_{=\xi}$$

where

$$\xi = \sum_i \sum_j W_n(i)W_n(j)\, E\{w(t-i)w(t-j)\}$$

$$= \sum_i \sum_j W_n(i)W_n(j)\, \sigma^2 \delta(i-j) = \sigma^2 \sum_i W_n^2(i)$$

Assuming mean ergodicity, the statistical mean is the same as the time average. To calculate $\gamma$, we replace the statistical average with the time average and approximate the function $f(t-i)$ with the first $n+2$ terms of its Taylor expansion around $i$ by assuming that the signal $f(t)$ is sufficiently smooth so that it has $n + 2$ continuous derivatives.

$$\gamma = \frac{1}{L}\sum_{t=1}^{L}\left(\sum_{i} W_n(i) f(t-i)\right)^2$$

$$\simeq \frac{1}{L}\sum_{t=1}^{L}\left(\sum_{i=-M}^{M} W_n(i)\left[f(t) + \frac{f^{(2)}(t)i^2}{2!} + \cdots + \frac{f^{(n+2)}(t)i^{n+2}}{(n+2)!}\right]\right)^2$$

$$= \frac{1}{L}\sum_{t=1}^{L}\left(\sum_{i=-M}^{M} W_n(i)\left[\frac{f^{(n+2)}(t)i^{n+2}}{(n+2)!}\right]\right)^2$$

(9)

where $f^{(n)}(t)$ is the $n$th order derivative of $f(t)$ with respect to $t$ and the last equality is due to lemma1.

the second term of equation can be calculated similarly. the result is

$$E\{y(t)\,f(t)\} == \frac{1}{L}\sum_{t} f^2(t) + \frac{1}{L}\sum_{t} f(t) \sum_{i=-M}^{M} W_n(i)\left[\frac{f^{(n+2)}(t)i^{n+2}}{(n+2)!}\right]$$

(10)

For the third term of equation (8), we have

$$E\{f^2(t)\} = \frac{1}{L}\sum_{t=1}^{L} f^2(t) \qquad (11)$$

Now, the final form of the cost function can be written using equations (8),(9), (10) and (11).

$$cost = MSE \simeq \frac{1}{L}\left(\sum_{t}(f^{(n+2)}(t)\,\mu)^2\right) + \sigma^2 \sum_{i} W_n^2(i) \quad (12)$$

where $\mu = \sum_{i=-M}^{M} W_n(i)\frac{i^{n+2}}{(n+2)!}$.

To calculate the cost function, we must obtain the terms $\mu$ and $\sum_i W_n^2(i)$ in terms of $M$ (or equivalently $N$). To calculate these terms, lemma 2 and 3 are expressed.

**Lemma2**: If $q_n'(x)$ is the first-order derivative of $q_n(x)$, then

$$\sum_{x=-M}^{M} W_n^2(x) = W_n(0) = \alpha_{n+1} q_{n+1}'(0)$$

**Proof**

Due to the equation $W_n(x) = \alpha_{n+1} \frac{q_{n+1}(x)}{x}$, it is sufficient to prove the following statement

$$c_{n+1} = \sum_x \left(\frac{q_{n+1}(x)}{x}\right)^2 = \frac{\frac{q_{n+1}(x)}{x}\bigg|_{x=0}}{\alpha_{n+1}} = \frac{q_{n+1}'(0)}{\alpha_{n+1}}$$

We first rewrite the recursive relation between polynomials.

$$(n+1)q_{n+1}(x) = 2(2n+1)x q_n(x) - n(N^2 - n^2) q_{n-1}(x)$$

After dividing both sides of this equation by $x$, raising to the power of 2 and getting the sum, we have

$$(n+1)^2 \sum_x \left(\frac{q_{n+1}(x)}{x}\right)^2 = (n+1)^2 c_{n+1}$$

$$= (2(2n+1))^2 \left(\sum_x q_n^2(x)\right)$$

$$- 4(2n+1)n(N^2 - n^2)\left(\sum_x \frac{q_{n-1}(x)}{x} q_n(x)\right)$$

$$+ n^2(N^2 - n^2)^2 c_{n-1} \quad (13)$$

The order of polynomial $\frac{q_{n-1}(x)}{x}$ is $n-2$ and it is perpendicular to $q_n(x)$, so $\sum_x \frac{q_{n-1}(x)}{x} q_n(x) = 0$. Now by multiplying both sides of the equation (13) by

$\alpha_{n+1}$, using equation (5) and the fact that $\binom{n}{\frac{n}{2}} = \frac{4(n-1)}{n}\binom{n-2}{\frac{n-2}{2}}$, after some straight forward calculations we have

$(n+1)\alpha_{n+1}c_{n+1}$
$$= 2(2n+1)\frac{(-1)^{\frac{n}{2}}}{2^n}(N^2-1^2)(N^2-3^2)\ldots(N^2-(n-1)^2)$$
$$- n(N^2-n^2)\alpha_{n-1}c_{n-1} \quad *$$

On the other hand, by calculating derivatives of the two sides of the recursive relation (6) and observing the result at $x = 0$, we have

$$(n+1)q'_{n+1}(0) = 2(2n+1)q_n(0) - n(N^2-n^2)q'_{n-1}(0) \quad **$$

According to equation (7) for $q_n(0)$ and comparing (*) and (**), we can conclude that

$$q'_{n+1}(0) = \alpha_{n+1}c_{n+1}$$

therefore

$$\sum_{x=-M}^{M} W_n^2(x) = W_n(0) = \alpha_{n+1}q'_{n+1}(0) \quad \blacksquare$$

Now according to equation (8), we must calculate $W_n(0)$. Lemma 3 determines its value versus filter order.

**Lemma3:**

For $N \gg 1$, the value of $W_n(0)$ is approximately equal to:

$$W_n(0) = \frac{1}{N}\left[\frac{(n+1)}{2^n}\binom{n}{\frac{n}{2}}\right]^2 = \frac{\beta_n}{N}$$

**Proof**

We use mathematical induction to prove this lemma. The statement holds for $n = 0$

$$W_0(0) = \frac{1}{N}$$

Assume that the statement holds for $n = k - 2$, we want to examine the statement correctness for $n = k$. At first, both sides of recursive relation (6) are multiplied by $\alpha_{k+1}$, so

$$(k+1)\alpha_{k+1}q'_{k+1}(0) = 2(2k+1)\alpha_{k+1}q_k(0) - k(N^2 - k^2)\left[\frac{\alpha_{k+1}}{\alpha_{k-1}}\right]\alpha_{k-1}q'_{k-1}(0) \quad (14)$$

Using lemma2, recursive relation (14) is simplified to

$$(k+1)W_k(0) = 2(2k+1)\alpha_{k+1}q_k(0) - k(N^2 - k^2)\left[\frac{-(k+1)}{(N^2 - k^2)k}\right]W_{k-2}(0)$$

considering (2) and (7), we have:

$$(k+1)W_k(0) = 2(2k+1)\frac{k+1}{2^{k+1}}\binom{k}{\frac{k}{2}}\frac{(-1)^{\frac{k}{2}}}{N(N^2 - 2^2)(N^2 - 4^2)\dots(N^2 - k^2)} \cdot \frac{(-1)^{\frac{k}{2}}}{2^k}\binom{k}{\frac{k}{2}}[(N^2 - 1^2)(N^2 - 3^2)\dots(N^2 - (k-1)^2)] + \frac{(k+1)}{N}\left[\frac{(k-1)}{2^{k-2}}\binom{k-2}{\frac{k}{2}-1}\right]^2$$

$$\simeq \frac{(k+1)}{N}\left[\frac{(k+1)}{2^k}\binom{k}{\frac{k}{2}}\right]^2$$

Where the last approximation results assuming $N \gg 1$. As a result, statement holds for $n = k$ so the induction is complete and lemma3 is proved. ∎

The only remaining term for determining the cost function is $\mu$ which is derived as follows

$$\mu = \sum_{i=-k}^{k} w_i \frac{i^{n+2}}{(n+2)!} = \frac{\alpha_{n+1}}{(n+2)!}\sum_x q_{n+1}(x)x^{n+1} = \frac{\alpha_{n+1}}{(n+2)!}u_{n+1}$$

where $u_n = \sum_x q_n(x)x^n$. Lemma 4 tells how to calculate $u_n$.

**Lemma 4**:

If $q_n(x)$ is a Chebyshev polynomial with order $n$, then

$$u_n = \sum_x q_n(x) x^n = N \frac{(n!)^2}{(2n+1)!} \prod_{k=1}^{n} (N^2 - k^2)$$

**Proof**

To obtain $u_n$, we use the recursive relation. Both sides of (6) are multiplied by $x^{n-1}$ and summed for all $x$. As a result we have

$$(n+1) \sum_x q_{n+1}(x) x^{n-1} - 2(2n+1) \sum_x q_n(x) x^n + n(N^2 - n^2) \sum_x q_{n-1}(x) x^{n-1} = 0$$

According to orthogonality of polynomials we have $\sum_x q_{n+1}(x) x^{n-1} = 0$ therefore

$$u_n = \frac{n(N^2 - n^2)}{2(2n+1)} u_{n-1}$$

Now we can write $u_{n-1}$ in terms of $u_{n-2}$ and so on $u_0$

$$u_n = \frac{n(n-1)(N^2 - n^2)(N^2 - (n-1)^2)}{2^2 (2n+1)(2n-1)} u_{n-2} = \cdots = \left[ \prod_{k=1}^{n} \frac{k(N^2 - k^2)}{2(2k+1)} \right] u_0$$

$$= \frac{N}{2^n} \prod_{k=1}^{n} \frac{k(N^2 - k^2)}{(2k+1)} = N \frac{(n!)^2}{(2n+1)!} \prod_{k=1}^{n} (N^2 - k^2)$$

where the last equation is according to the equation $\prod_{k=1}^{n} \frac{1}{(2k+1)} = \frac{n! \, 2^n}{(2n+1)!}$ ∎

Now by using lemma 4, we can obtain the expression for $\mu$ as follows:

$$\mu = \sum_{i=-k}^{k} w_i \frac{i^{n+2}}{(n+2)!} = \frac{\alpha_{n+1}}{(n+2)} \frac{N(n+1)!}{(2n+3)!} \prod_{k=1}^{n+1} (N^2 - k^2)$$
$$= h_n (N^2 - 1^2)(N^2 - 3^2) \ldots (N^2 - (n+1)^2)$$

where $h_n = \frac{(-1)^{\frac{n}{2}} (n+1)(n+1)!}{2^{n+1} (n+2)(2n+3)!} \binom{n}{\frac{n}{2}}$.

By assuming $N \gg n$ the statement for $\mu$ is simplified to

$$\mu \simeq h_n N^{n+2}$$

**Theorem 1**: consider the main signal $f(t)$ corrupted by iid noise $w(t)$ with zero mean and variance $\sigma^2$. The Savitzky-Golay filter with order $n$ and window length $N$ is used to reconstruct the main signal from its noisy version. Assuming $N \gg n$ and that the signal $f(t)$ is sufficiently smooth to have $n+2$ continuous derivatives, optimum window length of SG filter is approximately equal to

$$N_{opt} = \sqrt[2n+5]{\frac{2(n+2)\left((2n+3)!\right)^2}{\left((n+1)!\right)^2} \frac{\sigma^2}{v_n}}$$

where $v_n = \frac{1}{L}\sum_t \left(f^{(n+2)}(t)\right)^2$.

**Proof**

According to lemma2, lemma3 and lemma4, the cost function (12) for $N \gg n$ can be written as follows

$$\text{cost} = \frac{1}{L}\sum_t \left(f^{(n+2)}(t)\mu\right)^2 + \sigma^2 \sum_{i=-k}^{k} w_i^2 = v_n \mu^2 + \sigma^2 w(0)$$

$$= v_n(h_n^2 N^{2n+4}) + \frac{\sigma^2 \beta_n}{N}$$

To obtain optimum window length, we differentiate the cost function with respect to $N$ and set the result to zero.

$$\frac{\partial \text{cost}}{\partial N} = v_n h_n^2 (2n+4) N^{2n+3} - \frac{\sigma^2 \beta_n}{N} = 0$$

therefore

$$N_{opt} = \sqrt[2n+5]{\frac{2(n+2)\left((2n+3)!\right)^2}{\left((n+1)!\right)^2} \frac{\sigma^2}{v_n}} \quad (14) \blacksquare$$

As can be seen, the optimal window length depends on the noise power (through the variance term), the number of samples, the shape of the signal

(through the denominator term) and the order of the filter ($n$). Since $v_n$ is a function of the signal, the ratio $\frac{v_n}{\sigma^2}$ is conceptually similar to SNR (signal to noise ratio). The lower the SNR, the larger the optimal window length will be. Now we can obtain the minimum of MSE (MMSE) by inserting $N_{opt}$ in equation ().

$$MMSE = \left(\frac{2n+5}{2n+4}\right)\left(\frac{n+1}{2^n}\binom{n}{\frac{n}{2}}\right)^2 \sqrt[2n+5]{\frac{((n+1)!)^2}{2(n+2)((2n+3)!)^2}} (\sigma^2)^{\left(\frac{2n+4}{2n+5}\right)} v_n^{\frac{1}{2n+5}}$$

$$= r_n \sigma^{2\left(\frac{2n+4}{2n+5}\right)} v_n^{\frac{1}{2n+5}} \quad (16)$$

We investigate the effect of the coefficient $r_n$ in MMSE. The curve of this parameter versus filter order $n$ is depicted in Fig1.

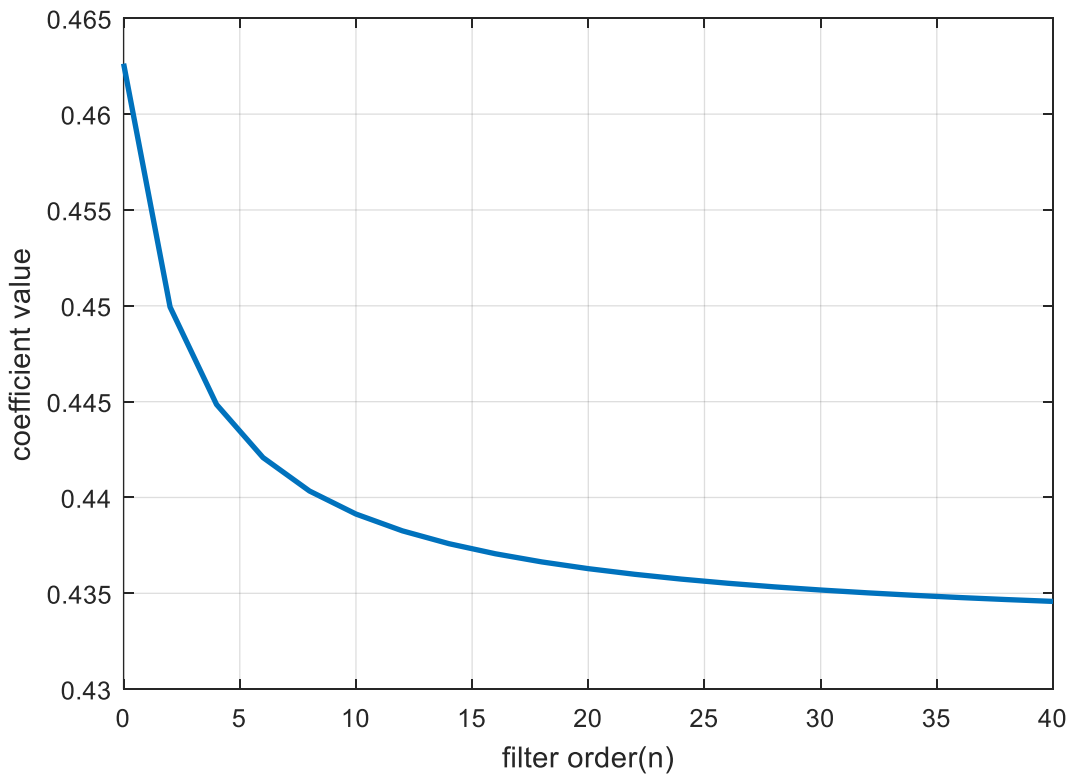

Fig 1: The effect of coefficient $r_n$ in MMSE (equation (16) ) versus filter order

As can be seen, the coefficient $r_n$ does not change much with the filter order and is approximately 0.45 for all values of $n$. We will investigate the effect of the filter order and the signal waveform in MMSE in section V.

To calculate optimum window length using (14), knowing $v_n$ (signal waveform) is necessary while being not available in most cases. To solve this problem, one can use the noisy signal $x(t)$ instead of the main signal $f(t)$. Based on this idea, we present an iterative method in algorithm1 for determining the optimal window length.

In this algorithm $SavGol(x, N, n)$ is a zero phase Savitzky-Golay filter with input $x$, window length $N$ and filter order $n$. The symbols $\lfloor \rfloor$ and $diff\{x, m\}$ denote the floor function and the $m$th-order derivative of input $x$, respectively. In calculating $v_n$, the first derivative of the smoothed noisy signal ($diff\{y, 1\}$) is smoothed by $SavGol$ in addition to the noisy signal ($x$), since the smoothed noisy signal is also noisy. The performance of this algorithm is investigated in the next section.

$$N_{opt} = 3; N_1 = 1;$$

$$\text{while } N_1 \neq N_{opt}$$

$$N_1 = 2\lfloor N_{opt}/2 \rfloor$$

$$y = SavGol(x, N_1, n)$$

$$dy = SavGol(diff(y, 1), N_1, n)$$

$$Y = diff(dy, 3)$$

$$c_1 = mean(Y^2)$$

$$N_{opt} = \sqrt[2n+5]{\frac{2(n+2)((2n+3)!)^2}{((n+1)!)^2} \frac{\sigma^2}{c_1}}$$

$$\text{end}$$

## V. Numerical results

In this section, first the effect of filter order and signal waveform is investigated, and then numerical results are compared with the theoretical ones. In all scenarios we assume that that noise power is known.

### Effect of the filter order

We want to choose a filter with appropriate parameters including filter order and window length which minimizes the MSE and has low complexity. Consider minimum MSE in equation (16). As previously mentioned, the coefficient $r_n$ dose not have much effect on the minimum MSE. The term $\sigma^{2\left(\frac{2n+4}{2n+5}\right)}$ indicates the effect of noise power and the filter order has little effect in this term of MMSE also. The term $v_n^{\frac{1}{2n+5}}$ depends on the signal waveform, number of samples and the filter order. As an example we consider three signal waveforms, each corrupted by Gaussian noise with zero mean and two values of $\sigma = 0.05$ (low noise power) and $\sigma = 1$ (high noise power). MMSE of each signal waveform with different filter order and noise power are presented in Table1. It can be seen that the minimum MSE and the optimal window length depend on the signal waveform.

Table1: The effect of waveform signal in the minimum MSE and the optimal window length

| $L = 1000, \quad T = 15$ | | $n = 0$ | | $n = 2$ | | $n = 4$ | | $n = 6$ | |
|---|---|---|---|---|---|---|---|---|---|
| | | MMSE | $N_{opt}$ | MMSE | $N_{opt}$ | MMSE | $N_{opt}$ | MMSE | $N_{opt}$ |
| $X_1 = 2\sin\left(\frac{2\pi t^2}{100}\right) + \cos\left(\frac{3\pi t}{100}\right)$ $0 \le t \le T$ | $\sigma = 0.05$ | 18.4e-5 | 17 | 7.7e-5 | 87 | 6.2e-5 | 171 | 4.9e-5 | 303 |
| | $\sigma = 1$ | 0.0256 | 59 | 0.0165 | 163 | 0.0135 | 321 | 0.0139 | 433 |
| $X_2 = 2\sin\left(\frac{2\pi t}{5}\right)$ $0 \le t \le T$ | $\sigma = 0.05$ | 5.2e-5 | 71 | 1.7e-5 | 217 | 1.5e-5 | 459 | 1.9e-5 | 468 |
| | $\sigma = 1$ | 0.0107 | 231 | 0.0055 | 393 | 0.0085 | 467 | 0.0118 | 471 |
| $X_3 = e^{\frac{t}{3}} + \sqrt{t}$ $0 \le t \le T$ | $\sigma = 0.05$ | 10.8e-5 | 33 | 4.2e-5 | 85 | 2.8e-5 | 207 | 2.4e-5 | 363 |
| | $\sigma = 1$ | 0.0265 | 57 | 0.0153 | 183 | 0.0130 | 343 | 0.0130 | 479 |

The purpose of increasing the filter order is to reduce the estimation error. As can be seen from the table, reduction of the minimum MSE from the filter order $n = 0$ (which is the moving average filter) to $n = 2$ is significant, but there is not much difference between order 2 and higher. On the other hand, increasing the filter order, in addition to increasing the computational burden caused by the LS fitting, increases the optimal window length, which in turn increases the amount of computational burden. Therefore, considering the performance and computational burden, selecting the filter order $n = 2$ is an appropriate choice. In the following we consider the filter order $n = 2$ and the waveform signal $X_1$.

**Effect of window length**

In this section the effect of window length in the time domain is intuitively investigated. Consider the signal $X_1$ corrupted by a noise with $\sigma = 1$. The noisy signal is shown by green dots in Fig.2. The noisy signal is filtered by the SG filter with order 2 and three different window lengths. The output of the filter for these window lengths are depicted in Fig.2. As can be seen, the filter output with a short window length ($N = 19$) has a low bias and high variance in comparison to the main signal ($f(t)$). As the window length increases, the variance decreases but the bias increases. In this tradeoff, there is an optimal point that has a good bias and variance. In this example this point is $N = 163$.

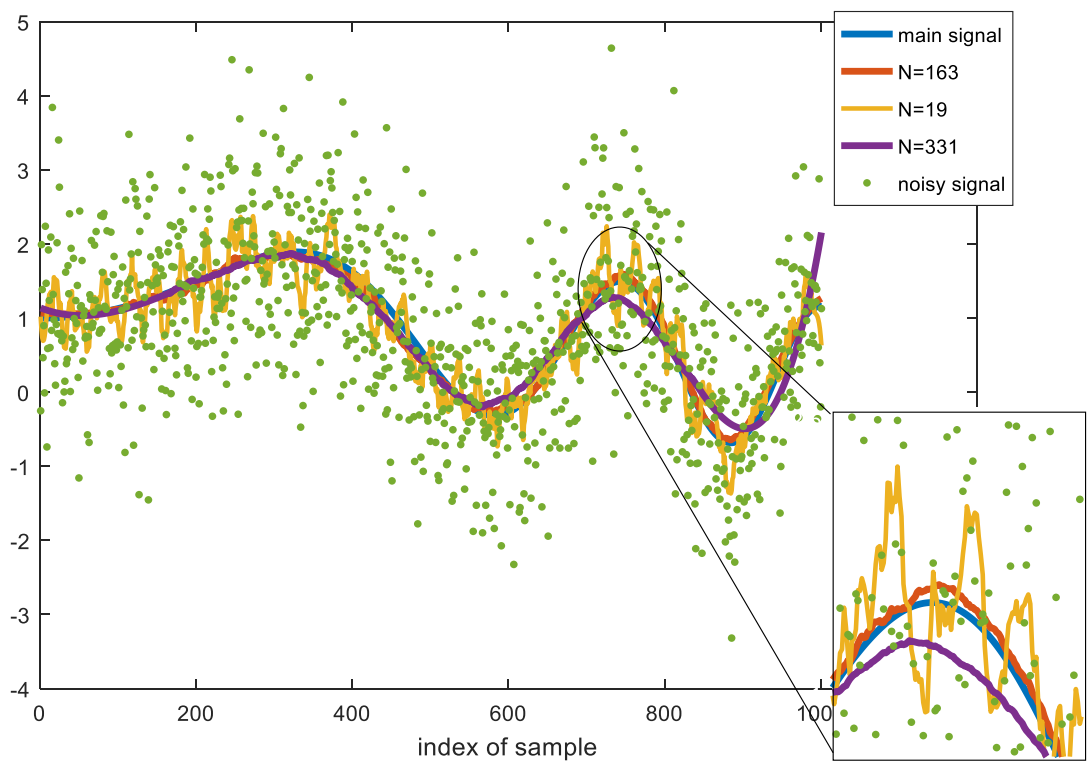

Fig 2: the bias-variance tradeoff and the effect of window length in reconstruction signal

**Algorithm evaluation**

In this section, we compare optimal window length obtained from algorithm1 with the actual value and the proposed method by Krishnan in [39]. Consider the signal $X_1$ corrupted by the noise with noise power $\sigma^2$. To evaluate algorithms, Monte Carlo technique with 100 iteration is used. The graph of the optimal window length versus power of noise is displayed in Fig.3. The higher the noise power, the longer the necessary window length will be, but this relation is non-linear. The proposed algorithm, in comparison with SURE method, has a better performance in estimating the optimal window length and is closer to the actual value. In addition, the proposed algorithm has a lower computational load compared to Krishnan method. For instance in this example, in Krishnan method, LS-function (SavGol) is called $\frac{L}{2} = 500$ times, but the algorithm1 converges at no more than 25 iterations and the max number of calls to LS-function is 50. Therefore, the proposed method has a better performance and can have less computational burden.

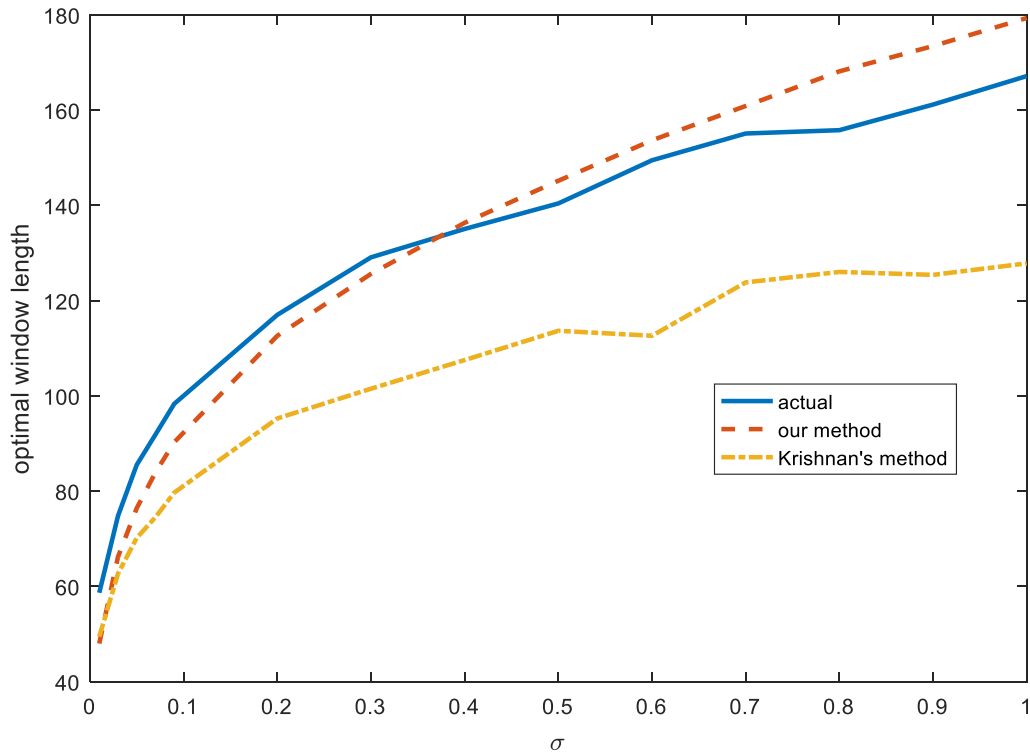

Fig. 3: Comparison of proposed method with Krishnan's method and actual results in calculating optimal window length

## VI.    Conclusion

In this paper, we investigated the problem of optimal window length determination for Savitzky-Golay filter. Based on the window length, there was a tradeoff between the bias and the variance of the estimation error. We provided a closed form formula for the optimal window length in the sense of minimizing MSE. The optimal window length depends on the noise power, number of samples, signal waveform and filter order. Calculating the optimal window length requires knowledge of the main signal. To solve this problem, an efficient algorithm with the help of a noisy signal was proposed which has better performance and lower computational load compared to existing methods.